\documentclass[twocolumn,superscriptaddress,showpacs,preprintnumbers,amsmath,amssymb,floatfix]{revtex4}
\usepackage{graphicx,dcolumn,bm}

\newcommand{\fig}[1]{Fig.\ \ref{#1}}

\newcommand{\etac}{ \eta_{\rm c} }
\newcommand{\etap}{ \eta_{\rm p} }

\newcommand{\kb}{ k_{\rm B} }
\newcommand{\Pc}{ P(\etac) }

\newcommand{\zc}{ z_{\rm c} }
\newcommand{\zp}{ z_{\rm p} }

\newcommand{\Nc}{ N_{\rm c} }
\newcommand{\Np}{ N_{\rm p} }

\newcommand{\sigmac}{ \sigma_{\rm c} }
\newcommand{\sigmap}{ \sigma_{\rm p} }
\newcommand{\sigmag}{ \sigma_{\rm G} }

\begin{document}

\title{Accurate description of bulk and interfacial properties in
colloid-polymer mixtures}

\author{R. L. C. Vink}
\affiliation{Institut f\"ur Physik, Johannes-Gutenberg-Universit\"at, 
Staudinger Weg 7, D-55099 Mainz, Germany}

\author{A. Jusufi}
\affiliation{Lehrstuhl f\"{u}r Physikalische Chemie I, Universit\"{a}t
Bayreuth, D-95440 Bayreuth, Germany}

\author{J. Dzubiella} 
\affiliation{NSF Center for Theoretical Biological Physics (CTBP),
Department of Chemistry and Biochemistry, University of California, San
Diego, La Jolla, California 92093-0365}

\author{C. N. Likos}
\affiliation{Institut f\"ur Theoretische Physik II,
Heinrich-Heine-Universit\"at D\"{u}sseldorf, D-40225 D\"{u}sseldorf, 
Germany}

\date{\today}

\begin{abstract}

Large-scale Monte Carlo simulations of a phase-separating colloid-polymer
mixture are performed and compared to recent experiments. The approach is
based on effective interaction potentials in which the central monomers of
self-avoiding polymer chains are used as effective coordinates. By
incorporating polymer nonideality together with soft colloid-polymer
repulsion, the predicted binodal is in excellent agreement with recent
experiments. In addition, the interfacial tension as well as the capillary
length are in quantitative agreement with experimental results obtained at
a number of points in the phase-coexistence region, without the use of any
fit parameters.

\end{abstract}



\maketitle


In addition to commercial applications, mixtures of colloids and
nonadsorbing polymer are interesting because of their analogy to atomic
systems \cite{poon:2004}. Much effort has been devoted to understand the
phase behavior of such mixtures. Of particular interest is phase
separation, which occurs when the polymer density and diameter of gyration
are sufficiently large, leading to the formation of two coexisting phases:
one phase lean in colloids and dense in polymers (the colloidal vapor) and
one phase dense in colloids and lean in polymers (the colloidal liquid).
As was shown by Asakura and Oosawa (AO), phase separation in
colloid-polymer mixtures (CPM) is driven by entropy
\cite{asakura.oosawa:1954}. In the AO description, colloids and polymers
are treated as effective spheres, assuming hard-sphere interactions
between colloid-colloid and colloid-polymer pairs, while the polymers can
interpenetrate freely. A major advance has been the development of a
geometry-based density functional for the AO-model \cite{schmidt:prl:00},
which has led to a host of novel and intriguing predictions regarding
interfacial phenomena within this model \cite{evans:review:03}. However,
when comparing to actual experiments, quantitative discrepancies arise. An
important deviation of the AO-model is that it underestimates the polymer
concentration in the colloidal liquid \cite{bolhuis.louis.ea:2002,
aarts.dullens.ea:2004}. This effect is more pronounced within the
free-volume approximation \cite{lek:92}, which is also the bulk limit of
the density functional of Ref.\ \cite{schmidt:prl:00}, and it persists
when the AO-binodals are obtained by means of computer simulations
\cite{bolhuis.louis.ea:2002, vink.horbach:2004*1}.

While the colloid-colloid interaction in realistic systems is indeed well
described by the hard sphere potential \cite{imhof.dhont:1995}, the
colloid-polymer and polymer-polymer interactions are more complex. To
circumvent the shortcomings of the AO-model, numerous different approaches
have been employed, both at the effective \cite{warren:pre:95,
schmidt.denton.ea:2003, vink.schmidt:2005, schmidt.fuchs:2002} and at the
monomer-resolved \cite{fuchs:pre:01,bryk:jcp:05} levels. As a general
trend, inclusion of polymer nonideality does improve on the major drawback
of the AO-model, i.e., it yields higher polymer concentrations in the
colloidal liquid \cite{schmidt.denton.ea:2003, schmidt.fuchs:2002}. A
remarkably accurate method to capture polymer nonideality is the ``polymer
as soft colloid'' approach \cite{louis.bolhuis.ea:2000}. Here, the
polymers' centers of mass are chosen as effective coordinates while all
fluctuating monomers are canonically integrated out. The sought-for
effective potentials are obtained by inverting the correlation functions
obtained in Monte Carlo (MC) simulations of self-avoiding random walks
(SAW) on a lattice \cite{louis.bolhuis.ea:2000,bolhuis.louis.ea:2001,
bolhuis.louis:macrom:02}. The pioneering work of Bolhuis {\it et al.}
\cite{bolhuis.louis:macrom:02} has led to the most accurate determination
of the phase behavior of CPM to date, demonstrated by direct comparison to
experiments \cite{bolhuis.louis.ea:2002}.

Despite the progress in calculating bulk phase diagrams, accurate
predictions of the interfacial tension $\gamma$ between coexisting phases,
a quantity that plays a key role in wetting and interfacial phenomena,
remain elusive. Theoretical approaches include the square-gradient
approximation \cite{aarts.dullens.ea:2004, brader:evans:epl:00,
moncho:jcp:03}, density-functional theory \cite{bryk:jcp:05,
brader:jpcm:02, moncho:jpcb:05} and simulations \cite{vink.horbach:2004*1,
vink.schmidt:2005, fortini.dijkstra.ea:2004}. Experimental data on
$\gamma$ are hard to obtain, mainly due to the very small value of this
quantity ($\gamma \sim 1\,\mu{\rm N}/{\rm m}$). In several cases
\cite{aarts.dullens.ea:2004,brader:evans:epl:00, brader:jpcm:02},
theoretical predictions have been compared to the experimental results of
Refs.\ \cite{dehoog:jpcb:99} and \cite{aarts:jpcm:03}.  These comparisons
are carried out by plotting the theoretical interfacial tension as a
function of the colloid density gap across the binodal, a procedure that
tends to 
obscure the fact that the theoretical and experimental binodals can be in
considerable disagreement \cite{aarts.dullens.ea:2004}. Recently,
Moncho-Jord{\`a} {\it et al.} \cite{moncho:jpcb:05} employed
density-functional theory to calculate the interfacial tension employing
the interactions of Ref.\ \cite{bolhuis.louis.ea:2002}. However, they
adopted a depletion picture, losing thereby the effect of polymer-induced
many-body interactions between the colloids. Thus, a full two-component
treatment of interacting CPM is necessary in order to capture bulk {\it
and} interfacial behavior quantitatively \cite{moncho:jpcb:05}.

In this Letter, we demonstrate that simulations of CPM using accurate
effective interactions predict bulk and interfacial properties correctly.
We consider the so-called ``colloid limit'', where the polymer diameter of
gyration $\sigmag$ is smaller than the colloid diameter $\sigmac$.
Recently, experimental measurements of the interfacial tension and
capillary length in the colloid limit became available, to which we can
compare \cite{aarts:2005*b}. At the same time, accurate effective
interactions for the colloid limit exist, obtained in off-lattice
molecular dynamics (MD) simulations of self-avoiding polymer chains
\cite{jusufi:etal:jpcm:01}. We choose here the central monomer to
represent the chain, canonically tracing out the remaining, fluctuating
monomers. If the coarse-graining procedure is carried out accurately, the
bulk thermodynamics of the mixture should be strictly {\it independent} of
the choice of the effective coordinates \cite{likos:pr:01}. Moreover, for
athermal solvents and in the limit of long chains, details of the
microscopic monomer-monomer interactions become irrelevant
\cite{degennes:book}. Thus, the phase behavior should be independent of
whether one adopts for the polymers a lattice, SAW-model as done in Ref.\
\cite{bolhuis.louis.ea:2002}, or the approach at hand. We will explicitly
check whether this requirement is fulfilled in what follows.

The colloids are represented by their centers and $r$ denotes the distance
between any effective coordinates in the corresponding effective
interactions $V_{ij}(r)$, with $i,j = {\rm c,p}$. The colloid-colloid
interaction $V_{\rm cc}(r)$ is given by a hard-sphere potential of
diameter $\sigmac$. The colloid-polymer interaction, $V_{\rm cp}(r)$
diverges for $r < \sigmac/2$ and for larger separations it reads as
\cite{jusufi:etal:jpcm:01}:
\begin{widetext}
\begin{eqnarray}
\nonumber
\label{eq:vcp}
  \beta V_{\rm cp}(r) &=& \frac{\sqrt{2}\Lambda\sigmac}{r}
    \left\{ \begin{array}{ll}
    \xi_2 - \ln \left( \frac{2r-\sigmac}{\sigmap} \right) -
      \left( \frac{(2r-\sigmac)^2}{\sigmap^2} - 1 \right)
      \left( \xi_1 - \frac{1}{2} \right)
      & {\rm if}\,\, \frac{\sigmac}{2} < r \leq \frac{\sigmac+\sigmap}{2}, \\
    \xi_2 \frac{ 1 - {\rm erf} ( \kappa (2r-\sigmac))}{ 1 - {\rm erf}
      (\kappa \sigmap)} & {\rm if}\,\, r > \frac{\sigmac+\sigmap}{2}
  \end{array} \right.
\end{eqnarray}
\end{widetext}
where $\sigmap$ a typical length scale given by $\sigmap = 0.66 \sigmag$,
$\beta = (\kb T)^{-1}$, $T$ the temperature, $\kb$ the Boltzmann constant,
$\xi_1 \equiv 1/(1 + 2 \kappa^2 \sigmap^2)$, $\xi_2 \equiv (\sqrt{\pi}
\xi_1 / \kappa \sigmap) [1 - {\rm erf} (\kappa \sigmap) ] \exp ( \kappa^2
\sigmap^2)$, and parameters $\Lambda$ and $\kappa$, determined in Ref.\
\cite{jusufi:etal:jpcm:01} by fitting to simulation results. The
corresponding polymer-polymer interaction is given by:
\begin{eqnarray}
\label{eq:vpp}
  \beta V_{\rm pp}(r) = 0.786\times 
  \left\{ \begin{array}{ll}
    -\ln \left( \frac{r}{\sigmap} \right) + \frac{1}{2 \tau^2
      \sigmap^2} & {\rm if}\,\, r \leq \sigmap, \\
    \frac{1}{2 \tau^2 \sigmap^2}
      \exp \left[ -\tau^2 (r^2 - \sigmap^2) \right]
      & {\rm if}\,\, r > \sigmap,
  \end{array} \right. \nonumber
\end{eqnarray}
with $\tau$ obtained by requiring that the effective interaction correctly
reproduces the second virial coefficient of dilute polymer solutions and
resulting in the value $\tau\sigmap=1.03$ \cite{jusufi:etal:jpcm:01}. To
match the experiment of Ref.\ \cite{aarts:2005*b}, we consider a mixture 
of colloids and polymers with size ratio $q \equiv \sigmag / \sigmac = 
0.56$. The parameters are $\Lambda=0.46$, and $\kappa \sigmap = 0.52715$
\cite{footnote}.


The binodal and the interfacial tension are obtained in the grand
canonical ensemble, by MC simulation of a mixture of $\Nc$ colloids and
$\Np$ polymers, interacting via the above pair potentials. In this
ensemble, the temperature, the volume $V$, and the respective fugacities,
$\zc$ and $\zp$, of colloids and polymers, are fixed, while the number of
particles in the system fluctuates. We also introduce the colloid and
polymer packing fractions $\eta_{\rm{c,p}} = (\pi\sigma_{\rm{c,G}}^3/6)
N_{\rm{c,p}}/V$. Since the interactions are athermal, the temperature
plays no role and the phase behavior is set by $q$ and the fugacities. The
polymer fugacity is used as control parameter, analogous to inverse
temperature in fluid-vapor transitions in atomic systems. For a given
$\zp$, we measure the distribution $\Pc$, defined as the probability of
observing a system with colloid packing fraction $\etac$. For $\zp$
sufficiently far away from the critical point, phase coexistence is
obtained by tuning $\zc$ so that $\Pc$ becomes bimodal, with two peaks of
equal area. The peak at low $\etac$ corresponds to the colloidal vapor
phase, the peak at high $\etac$ to the colloidal liquid, and the region in
between to phase-separated states \cite{vink.horbach:2004*1}. The average
peak locations yield the colloid packing fractions in the two phases. The
interfacial tension is obtained from the average height of the peaks
\cite{binder:1982, vink.horbach:2004*1}. In order to simulate efficiently,
a grand canonical cluster move is used \cite{vink.horbach:2004*1}, in
combination with a reweighting scheme \cite{virnau.muller:2004}.


\begin{figure}
\begin{center}
\includegraphics[clip=,width=7cm]{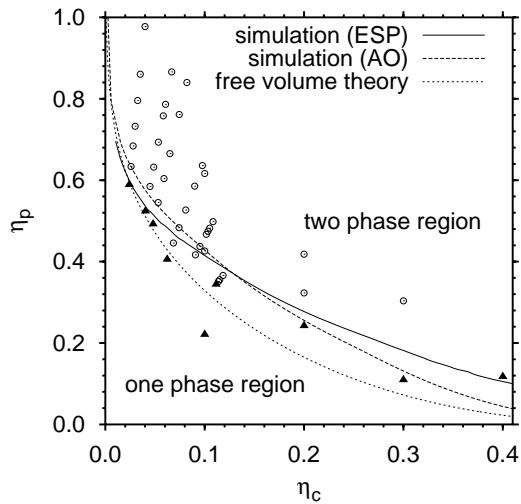}

\caption{\label{phasediag} Colloid-polymer binodals for $q=0.56$ obtained
in simulations using two different models (ESP and AO). Open circles are
experimental state-points at which phase separation was observed; black
triangles are experimental state-points at which only one phase was
observed. The experimental data were taken from Refs.\ \cite{aarts:2005*b}
and \cite{ilett.orrock.ea:1995}. The free-volume AO-binodal is also
shown.}

\end{center} 
\end{figure}

To obtain the binodal, we vary $\zp$ and record the corresponding
densities of colloids and polymers in the two coexisting phases. This
yields the phase diagram in system representation, which may directly be
compared to experiments, see \fig{phasediag}. The solid curve is the
binodal of the present work incorporating the effective soft potentials
(ESP)  $V_{\rm cp}(r)$ and $V_{\rm pp}(r)$ above. For comparison, the
dashed line shows the binodal of the AO model with $q=0.56$, obtained
following the same grand canonical simulation procedure. The free volume
result is also shown. As seen in \fig{phasediag}, ESP interactions give an
accurate description of the experimental binodal. Note in particular the
significant increase in the polymer density at high colloid density in
comparison to the AO-result.


\begin{figure}
\begin{center}
\includegraphics[clip=,width=7cm]{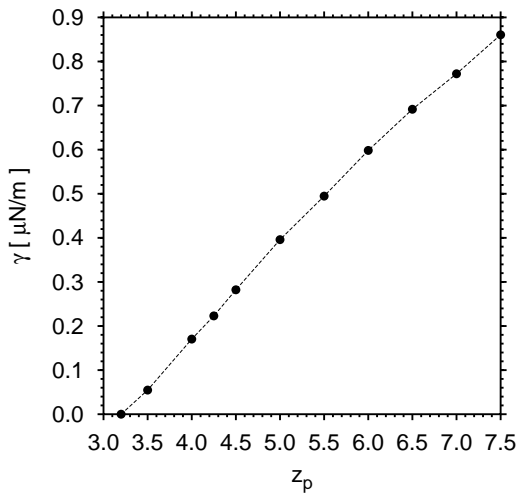}

\caption{\label{gamma} Interfacial tension $\gamma$ against the 
polymer fugacity from our ESP-simulations. The
line is a guide to the eye.}

\end{center}
\end{figure}

\begin{figure}
\begin{center}
\includegraphics[clip=,width=7cm]{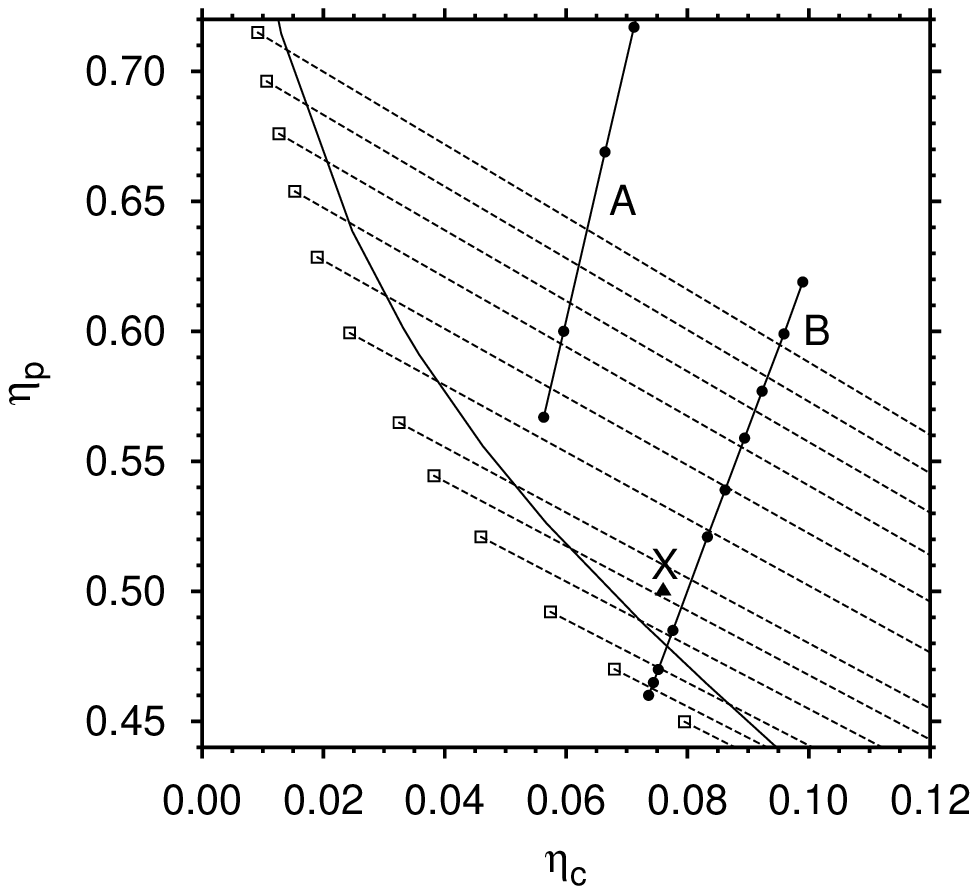}

\caption{\label{tielines} Close-up of the vapor branch of the binodal.  
Open squares with tielines (dashed) are binodal points obtained using ESP
interactions. The polymer fugacity corresponding to the tielines reads
$\zp = 7.5 \to 7 \to 6.5 \to 6 \to 5.5 \to 5 \to 4.5 \to 4.25 \to 4 \to
3.75 \to 3.5 \to 3.45$ (from top to bottom). The solid curve is the
binodal of the AO model. The X-symbol ($\etac=0.076$,
$\etap=0.50$) marks a state-point at which the interfacial tension was
measured experimentally. The closed circles on the dilution lines~A and~B
are experimental state-points at which the capillary length was
measured.}
 
\end{center}
\end{figure}

Next, we consider the interfacial tension, and compare to the recent
experiment of Ref.\ \cite{aarts:2005*b}. The colloids are PMMA spheres
($\sigmac=50\,{\rm nm}$), mixed with polymer with diameter of gyration
$\sigmag=28\,{\rm nm}$, and dissolved in decalin at $T=298\,{\rm K}$. In
\fig{gamma}, we show simulation results of the interfacial tension as
function of $\zp$. As expected, the tension decreases markedly upon
lowering $\zp$, vanishing at the critical point (the critical polymer
fugacity is approximately $z_{\rm p,cr} \approx 3.2$). To enable the
comparison to experiment, we show in \fig{tielines} several binodal
tielines obtained in the simulation. At point~X, which is close to the
tieline corresponding to $\zp=4.25$, the experimental interfacial tension
equals $\gamma=0.16-0.2$~$\mu$N/m \cite{aarts:2005*b}. The corresponding
interfacial tension in the simulation reads $\gamma=0.22$~$\mu$N/m, which
exceeds the experiment by only 10\%. In contrast, the interfacial tension
obtained {\it at the same state-point} in a {\it simulation} of the AO
model is $\gamma=0.07$~$\mu$N/m, which underestimates the experiment by
over 50\%. This is due to subtle differences in the location and range of
the critical regions of the two models. More precisely, defining the
distance from the critical point as $t \equiv \zp/z_{\rm p,cr}-1$, we
obtain for point~X using ESP interactions $t \approx 0.33$, but only $t
\approx 0.05$ using the AO model. For the AO model, point~X is thus much
closer to criticality and hence the interfacial tension is lower. At the
same time, if one calculates the interfacial tension for the AO-model
using the {\it free volume} tieline with the same colloid density gap as
in the experiment, the value $\gamma=0.5\,\mu{\rm N/m}$ is obtained
\cite{aarts:2005*b}. These discrepancies within the AO-model demonstrate
the large effect that inaccuracies in the binodals have on $\gamma$, as
well as ambiguities that arise by comparing interfacial tensions at
``rescaled'' state-points. In our approach, on the contrary, we provide a
comparison to the experimentally measured interfacial tension {\it in
absolute terms}, i.e., at the same state-point $(\eta_{\rm c}, \eta_{\rm
p})$.


\begin{figure}
\begin{center}
\includegraphics[clip=,width=7cm]{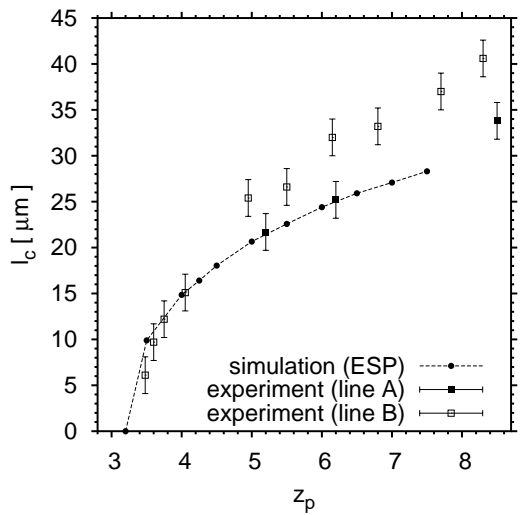}

\caption{\label{cap} Capillary length as function of the polymer fugacity
obtained in simulations (ESP), as well as in experiments along dilution
lines~A and~B of \fig{tielines}. The line connecting the simulation data
serves as a guide to the eye.}

\end{center}
\end{figure}

Finally, we compare to experimental measurements of the capillary length
$l_{\rm c} = \sqrt{\gamma / (g \Delta\varrho)}$, with $g=9.81$ m/s$^2$ the
gravitational acceleration, interfacial tension $\gamma$, and
$\Delta\varrho$ the mass density difference between the colloidal liquid
and vapor phase. For two coexisting phases with colloid- and
polymer-packing fraction gaps $\Delta\etac$ and $\Delta\etap$
respectively, it holds $\Delta\varrho = \Delta\etac(\varrho_{\rm c} -
\varrho_{\rm d}) + \Delta\etap m_{\rm p}/v_{\rm p}$, with the mass
densities $\varrho_{\rm c} = 1170$~kg/m$^3$ for PMMA and $\varrho_{\rm d}
= 890$~kg/m$^3$ for decalin, $m_{\rm p} = 3.87 \times 10^{-22}$~kg the
single polymer mass, and $v_{\rm p} = \pi\sigmag^3/6$ the effective single
polymer volume \cite{aarts:2005*b}. In \fig{cap}, we plot the capillary
length as obtained in the simulation as function of $\zp$. The
experimental data are also shown, where the conversion to $\zp$ was
performed with the aid of \fig{tielines} (experimental measurements at
$\zp>7.5$ were converted using linear extrapolation). As $\gamma =
g\Delta\varrho l_{\rm c}^2$ and since $\Delta\varrho$ is given accurately
by our approach (as shown by the good agreement with the experimental
binodal), the agreement with experiment regarding $l_{\rm c}$ directly
implies agreement with the interfacial tension $\gamma$ at all considered
state-points. Another striking feature of \fig{cap} is the remarkably good
agreement between simulation and experiment close to the critical point.
For 3D Ising systems, the capillary length is expected to vanish at
criticality as $l_{\rm c} \sim t^{\lambda}$ with the critical exponent
$\lambda = \nu-\beta/2 \approx 0.47$. While the present simulation and
experimental data are not accurate enough to extract the exponent, the
{\em curvature} of the data is certainly compatible with the anticipated
exponent. In particular, the data seem to approach the $\zp$ axis with
perpendicular slope.

In summary, we have demonstrated that the ``polymer as soft colloid''
approach \cite{louis.bolhuis.ea:2000}, using accurate effective
colloid-polymer and polymer-polymer interactions, not only reproduces the
experimental binodal, but also the interfacial tension and the capillary
length. Note that excellent agreement with the experimental binodal was
also obtained in Ref.\ \cite{bolhuis.louis.ea:2002}, in which the polymer
centers-of-mass were employed as effective coordinates. Both the effective
description of Ref.\ \cite{bolhuis.louis.ea:2002}, and the one adopted
here, thus reproduce the correct thermodynamics, providing a strong
confirmation of the power and self-consistency of coarse-graining
techniques. We anticipate that a full, two-component calculation of the
interfacial tension using the effective interactions of Ref.\
\cite{bolhuis.louis.ea:2002} will also capture the interfacial properties
correctly; the latter should be the subject of further investigations.
Additional experimental work in measuring interfacial tensions in
colloid-polymer mixtures is also highly desirable.

We thank Dirk Aarts for helpful discussions. This work was supported by
the DFG through the SFB-TR6. Allocation of computer time on the JUMP at
the Forschungszentrum J\"ulich is gratefully acknowledged.


\begin{thebibliography}{99}

\bibitem{poon:2004} W. Poon, Science {\bf 304}, 830 (2004).

\bibitem{asakura.oosawa:1954} S. Asakura and F. Oosawa, J. Chem. Phys.
{\bf 22}, 1255 (1954).

\bibitem{schmidt:prl:00} M. Schmidt {\it et al.}, Phys. Rev. Lett. {\bf
85}, 1934 (2000).

\bibitem{evans:review:03} J. M. Brader {\it et al.}, Mol. Phys. {\bf 101},
3349 (2003).

\bibitem{bolhuis.louis.ea:2002} P. G. Bolhuis {\it et al.}, Phys. Rev.
Lett. {\bf 89}, 128302 (2002).

\bibitem{aarts.dullens.ea:2004} D. G. A. L. Aarts {\it et al.}, J. Chem.
Phys. {\bf 120}, 1973 (2004).

\bibitem{lek:92} H. N. W. Lekkerkerker {\it et al.}, Europhys. Lett. {\bf
20}, 559 (1992).

\bibitem{vink.horbach:2004*1} R. L. C. Vink and J. Horbach, J. Chem. Phys.
{\bf 121}, 3253 (2004).

\bibitem{imhof.dhont:1995} A. Imhof and J. K. G. Dhont, Phys. Rev. Lett.
{\bf 75}, 1662 (1995).

\bibitem{warren:pre:95} P. B. Warren {\it et al.}, Phys. Rev. E {\bf 52},
5205 (1995).

\bibitem{schmidt.denton.ea:2003} M. Schmidt {\it et al.}, J. Chem. Phys.
{\bf 118}, 1541 (2003).

\bibitem{vink.schmidt:2005} R. L. C. Vink and M. Schmidt,
cond-mat/0501037.

\bibitem{schmidt.fuchs:2002} M. Schmidt and M. Fuchs, J. Chem. Phys. {\bf
117}, 6308 (2002).

\bibitem{fuchs:pre:01} M. Fuchs and K. S. Schweizer, Phys. Rev. E {\bf
64}, 021514 (2001);  J. Phys.: Condens. Matter {\bf 14}, R239 (2002).

\bibitem{bryk:jcp:05} P. Bryk, J. Chem. Phys. {\bf 122}, 064902 (2005).

\bibitem{louis.bolhuis.ea:2000} A. A. Louis {\it et al.}, Phys. Rev. Lett.
{\bf 85}, 2522 (2000).

\bibitem{bolhuis.louis.ea:2001} P. G. Bolhuis {\it et al.}, J. Chem. Phys.
{\bf 114}, 4296 (2001).

\bibitem{bolhuis.louis:macrom:02} P. G. Bolhuis and A. A. Louis,
Macromolecules {\bf 35}, 1860 (2002).

\bibitem{brader:evans:epl:00} J. M. Brader and R. Evans, Europhys. Lett.
{\bf 49}, 678 (2000).

\bibitem{moncho:jcp:03} A. Moncho-Jord{\`a} {\it et al.}, J. Chem. Phys.
{\bf 119}, 12667 (2003).

\bibitem{brader:jpcm:02} J. M. Brader {\it et al.}, J. Phys.: Condens.
Matter {\bf 14}, L1 (2002).

\bibitem{moncho:jpcb:05} A. Moncho-Jord{\`a} {\it et al.}, J. Phys. Chem.
B {\bf 109}, 6640 (2005).

\bibitem{fortini.dijkstra.ea:2004} A. Fortini {\it et al.},
cond-mat/0501134.

\bibitem{dehoog:jpcb:99} E. H. A. de Hoog and H. N. W. Lekkerkerker, J.
Phys. Chem. B {\bf 103}, 5274 (1999).

\bibitem{aarts:jpcm:03} D. G. A. L. Aarts {\it et al.}, J. Phys.: Condens.
Matter {\bf 15}, S245 (2003).

\bibitem{aarts:2005*b} D. G. A. L. Aarts, J. Phys. Chem. B {\bf 109}, 7407
(2005).

\bibitem{jusufi:etal:jpcm:01} A. Jusufi {\it et al.}, J. Phys.: Condens.
Matter {\bf 13}, 6177 (2001).

\bibitem{likos:pr:01} C. N. Likos, Phys. Rep. {\bf 348}, 267 (2001).

\bibitem{degennes:book} P.~G. de Gennes, {\it Scaling Concepts in Polymer
Physics}, Cornell University Press (Ithaca, N.Y., 1979).

\bibitem{footnote} The value of $\kappa \sigmap$ quoted in Ref.\
\cite{jusufi:etal:jpcm:01} is slightly higher. Extensive additional MD
simulations revealed that a lower value of $\kappa\sigmap$ yields a better
fit to MD data.

\bibitem{binder:1982} K. Binder, Phys. Rev. A {\bf 25}, 1699 (1982).

\bibitem{virnau.muller:2004} P. Virnau and M. M{\"u}ller, J. Chem. Phys.
{\bf 120}, 10925 (2004).

\bibitem{ilett.orrock.ea:1995} S. M. Ilett {\it et al.}, Phys. Rev. E {\bf
51}, 1344 (1995).

\end{thebibliography}
\end{document}